\documentclass[a4paper,11pt]{article}
\pdfoutput=1 % if your are submitting a pdflatex (i.e. if you have
\usepackage{jinstpub} % for details on the use of the package, please
\title{Radiation Hardness tests with neutron flux on different Silicon photomultiplier devices}

\author[a,1]{P.W.~Cattaneo,\note{Corresponding author.}}
\author[a,b]{T.~Cervi,}
\author[a,b]{A.~Menegolli,}
\author[c]{M.~Oddone,}
\author[d]{M.~Prata,}
\author[a]{M.C.~Prata,}
\author[a]{M.~Rossella}

\affiliation[a]{Istituto Nazionale di Fisica Nucleare (INFN) - Sezione di Pavia,\\Via Bassi 6, 27100 Pavia, Italy}
\affiliation[b]{Universit\`a degli Studi di Pavia - Dipartimento di Fisica,\\Via Bassi 6, 27100 Pavia, Italy}
\affiliation[c]{Universit\`a degli Studi di Pavia - Dipartimento di Chimica,\\Viale Taramelli 12, 27100 Pavia, Italy}
\affiliation[d]{Universit\`a degli Studi di Pavia - Laboratorio Energia Nucleare Applicata,\\Via Aselli 41, 27100 Pavia, Italy}

\emailAdd{Paolo.Cattaneo@pv.infn.it}

\abstract{
Radiation hardness is an important requirement for solid state readout devices operating in high radiation environments common in particle physics experiments. The MEG~II experiment, at PSI, Switzerland, investigates the forbidden decay $\mu^+ \to \mathrm{e}^+ \gamma$. Exploiting the most intense muon beam of the world. A significant flux of non-thermal neutrons (kinetic energy $E_k\geq 0.5$~MeV) is present in the experimental hall produced along the beam-line and in the hall itself. We present the effects of neutron fluxes comparable to the MEG~II expected doses on several Silicon Photomultiplier (SiPMs). The tested models are: AdvanSiD ASD-NUV3S-P50 (used in MEG~II experiment), AdvanSiD ASD-NUV3S-P40, AdvanSiD ASD-RGB3S-P40, Hamamatsu and Excelitas C30742-33-050-X. The neutron source is the thermal Sub-critical Multiplication complex (SM1) moderated with water, located at the University of Pavia (Italy). We report the change of SiPMs most important electric parameters: dark current, dark pulse frequency, gain, direct bias resistance, as a function of the integrated neutron fluency.
}
\proceeding{Instrumentation for colliding Beam Physics (INSTR-17)\\
  when 27-02-2017/03-03-2017\\
  where Novosibirsk (RU)}

\begin{document}
\maketitle
\flushbottom
\section{The MEG experiment and the MEG~II upgrade}
\label{sec:MEG}
The MEG experiment \cite{Adam:2013vqa} has been operational in the years 2008-2013 at the Paul Scherrer Institute (Villigen, CH), looking for the lepton flavor violating decay 
$\mu^+ \to e^++\gamma$. This process is highly suppressed in the Standard Model (SM) (branching ratio $BR < 5\times 10^{-50}$). 
Nevertheless, a measurable branching ratio is anticipated by many SM extensions \cite{barbieri,hisano-1999,calibbi}.

Detection of such a decay would be an unambiguous signal of physics beyond the SM, while improving its upper limit would constraint new theories. The kinematics of the signal consists in the 
two-body decay of a particle at rest: a positron and a photon with the same energies ($52.8$~MeV, half of the muon mass) emitted in time coincidence with opposite directions.
A precise measurement of the positron timing with a Timing Counter (TC) is crucial to discriminate between signal and combinatorial background from separate muon decays. 
In MEG the Timing Counter consisted of two sets of scintillator bars read-out by photomultipliers \cite{megtc}.

The MEG~II experiment is a project for upgrading MEG and improve its sensitivity of an additional order of magnitude \cite{Baldini:2013ke}. 
In MEG~II the role of the TC is taken by a pixelated Timing Counter (pTC) \cite{megtc1,megtc2}.
It consists of two arrays of thin scintillator plates readout by SiPMs located symmetrically to the decay target. A large number of SiPMs (6144) are employed to read out the scintillating 
light from plastic scintillator pixels designed to measure the time of arrival of positrons. In MEG~II we expect a flux of non-thermal neutron of $\sim 10^8$ n~ cm$^{-2}$ 
during the lifetime of the experiment due mainly to production along the beam with a kinetic energy distributed at $E_k > 0.5$~MeV. Those neutrons can damage the semiconductor 
devices located inside the experimental area.

\section{Radiation hardness tests with neutron flux}
\subsection{Tested SiPM models}
We irradiated different SiPM models: the AdvanSiD ASD-NUV3S-P50 (used in MEG~II
experiment), the AdvanSiD ASD-NUV3S-P40 and ASD-RGB3S-P40, the Hamamatsu S12572-050P and the Excelitas C30742-33-050-X. The characteristics of those devices are summarised in Table~\ref{tab:sipms}.
\begin{table}[htbp]
\centering
\caption{\label{tab:sipms} Characteristics of SiPMs under test.}
\smallskip
\begin{tabular}{|l|c|c|c|c|c|}
\hline
 & ASD-NUV3S & ASD-NUV3S& ASD-RGB3S&S12572 & C30742 \\
  & -P50 & -P40& -P40&-050P & -33-050-X \\
\hline
Active Area& $3\times 3~\mathrm{mm}^2$ &$3\times 3~\mathrm{mm}^2$ &$3\times 3~\mathrm{mm}^2$ &$3\times 3~\mathrm{mm}^2$ &$3\times 3~\mathrm{mm}^2$ \\
Pixel size& $50~\mu$m &$40~\mu$m &$40~\mu$m &$50~\mu$m  &$50~\mu$m \\
Number of Pixels& 3600 &5200 &5200  &3600 &3600\\
Fill Factor&  &60\%  &60\% &62\% &\\
Dark Counts& $1200~\mathrm{kcps}^{(1)}$   &$900~\mathrm{kcps}$ &$1800~\mathrm{kcps}$ &$1000~\mathrm{kcps}$ &$1350~\mathrm{kcps}$\\
$V_{bd}~^{(2)}$  &$24\pm 2~V$ &$24\pm 2~V$ &$25\pm 2~V$ &$65\pm 10~V$ &$95~V$\\
BVTC$^{(3)}$ & $26~~\mathrm{mV}/^{\circ}C$ & $26~~\mathrm{mV}/^{\circ}C$ &$26~~\mathrm{mV}/^{\circ}C$ &$60~~\mathrm{mV}/^{\circ}C$ & $90~~\mathrm{mV}/^{\circ}C$ \\ 
$\lambda_p~^{(4)} $ &420 nm & 420 nm & 550 nm &450 nm &$520 nm$\\
\hline
\end{tabular}
\begin{flushleft}
\footnotesize{(1). kcps = kilo counts per seconds.\\ }
\footnotesize{(2). $V_{bd}$ = Breakdown Voltage.\\ }
\footnotesize{(3). BVTC = Breakdown Voltage Temperature Coefficient.\\ }
\footnotesize{(4). $\lambda_P$ = Peak sensitivity wavelength. }
\end{flushleft}
\end{table}

\subsection{The SM1 facility}
\begin{figure}[htbp]
\centering % \begin{center}/\end{center} takes some additional vertical space
\includegraphics[width=\textwidth,trim=30 110 0 0,clip]{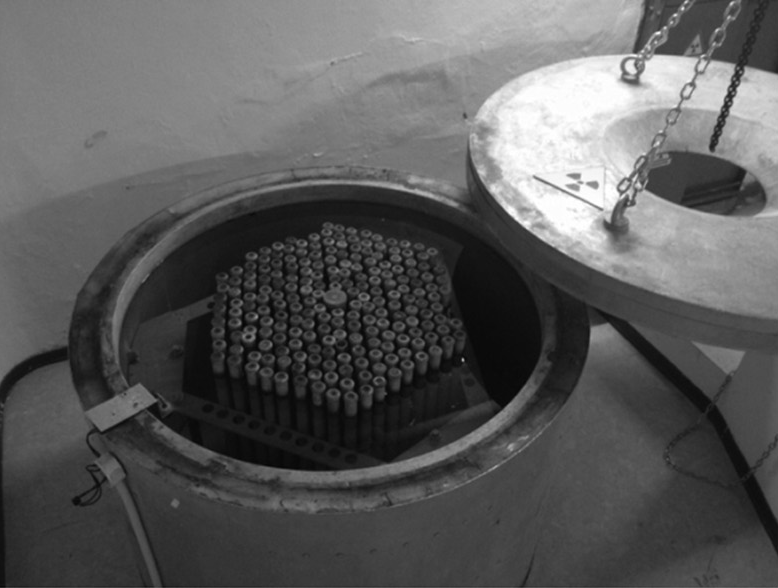}
\caption{\label{fig:SM1} SM1 - thermal Sub-critical Multiplication system.}
\end{figure}

SM1 is a thermal Sub-critical Multiplication complex moderated with water located at the Department of Chemistry,
University of Pavia (Italy) that is readily available for irradiation purposes \cite{odd:sm1}. 
The fuel is natural uranium in metallic form arranged in 206 Aluminum-clad fuel elements with an inner diameter of $2.8$~cm and a length of $132$~cm (see Fig.~\ref{fig:SM1}).

\begin{figure}[htbp]
\centering % \begin{center}/\end{center} takes some additional vertical space
\includegraphics[width=0.7\textwidth]{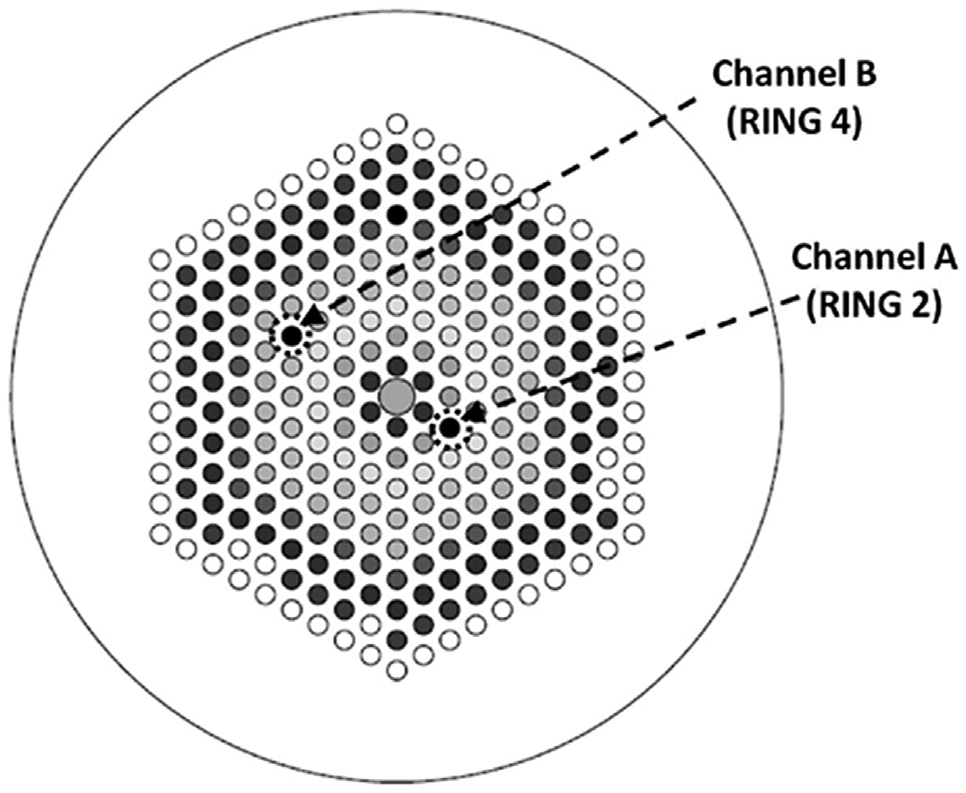}
\caption{\label{fig:fuel} Fuel elements placement in SM1.}
\end{figure}

\begin{figure}[htbp]
\centering % \begin{center}/\end{center} takes some additional vertical space
\includegraphics[width=\textwidth]{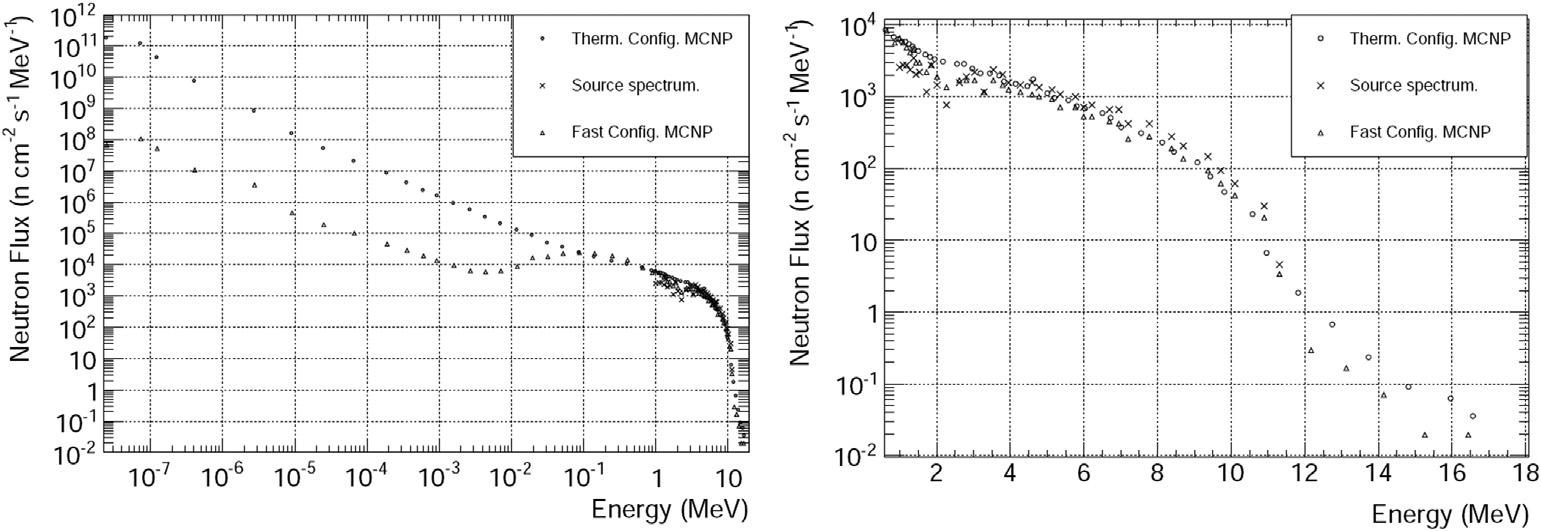}
\caption{\label{fig:spect} Comparison between the neutron flux spectrum obtained with MCNP~\textsuperscript{\textregistered} (Monte Carlo N-Particle) \cite{mcnp} simulations (circles) and experimental data processed with the SAND II code (crosses) for the irradiation channel A.}
%\caption{\label{fig:spect} Comparison between the neutron flux spectrum obtained with MCNP~\textsuperscript{\textregistered} \footnote{Monte Carlo N-Particle} \cite{mcnp} simulations (circles) and experimental data processed with the SAND II code (crosses) for the irradiation channel A.}
\end{figure}

Fuel elements are assembled in a hexagonal prism geometrical configuration  with a radial dimension of $114$~cm and a height of $135$~cm (see Fig.~\ref{fig:fuel}). 
Located at the centre of the SM1 core, a Pu-Be neutron source has an emission rate equal to $8.9\times 10^6~ \mathrm{s}^{-1}$ over the full solid angle \cite{pube}. 
Two channels are readily available for irradiation, in this paper we always use Channel A (Ring 2). The neutron spectra expected from Monte Carlo simulations 
in different configurations, thermal and fast, at channel A are shown in Fig.~\ref{fig:spect} compared with the experimental data processed with the SAND II code \cite{sandII}. 
The SAND II code is able to obtain neutron energy spectra by an analysis of experimental activation detector data.

The expected integrated neutron fluxes at position A in the thermal configuration, 
used in the irradiation, is $(5.9 \pm 0.2)\times 10^4$~n cm$^{-2}{\mathrm s}^{-1}$ (including slow and fast neutrons). In order to suppress the low energy neutrons below 0.5 eV, 
the devices were inserted in a Cd box 0.55 mm thick. We evaluated that, at position A, the neutron flux inside the box is $\sim 4 \times 10^4$~n cm$^{-2}{\mathrm s}^{-1}$.
Therefore the irradiation time required to deliver the total fluence expected in MEG~II is $\sim (3-5)\times  10^3$s.

The irradiation consists in a controlled exposition to the neutron flux for several seconds/minutes at the SM1 facility. All measurements on SiPMs have been done at fixed temperature of $30~^{\circ}$C. Every 1000~s of irradiation (integrated dose $\sim 4\times 10^7$~n cm$^{-2}$), each device has been characterised in term of I-V curve, breakdown voltage, noise. We have repeated this procedure several times up to a total exposition of $1\times 10^4$~s (integrated dose $\sim 4\times 10^8$~n cm$^{-2}$). In the following, we report the directly measured irradiation time rather than the dose deduced by simulation.

\section{Results}
\subsection{I-V curve, breakdown voltage and quenching resistance} 
The I-V curves have been measured using a Keithley Picoammeter/Voltage Source 6487 connected to a PC with an USB-GPIB converter (National Instruments model GPIB-USB-HS) and controlled with a Labview program.
Each device under test has been kept at constant temperature  ($30~^{\circ}C$), regulated by a Gefran Temperature Controller (model 1200). 

\begin{figure}[htbp]
\centering % \begin{center}/\end{center} takes some additional vertical space
\includegraphics[width=0.49\textwidth]{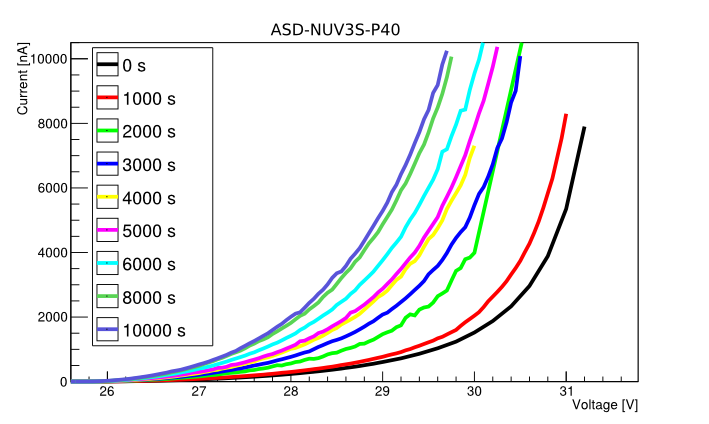}
\includegraphics[width=0.49\textwidth]{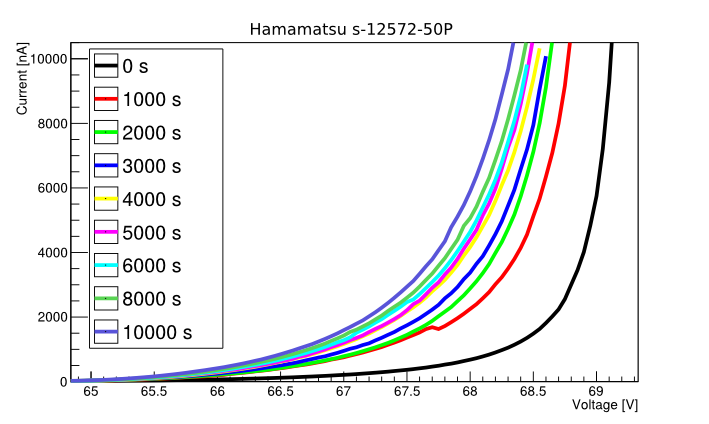}
\includegraphics[width=0.49\textwidth]{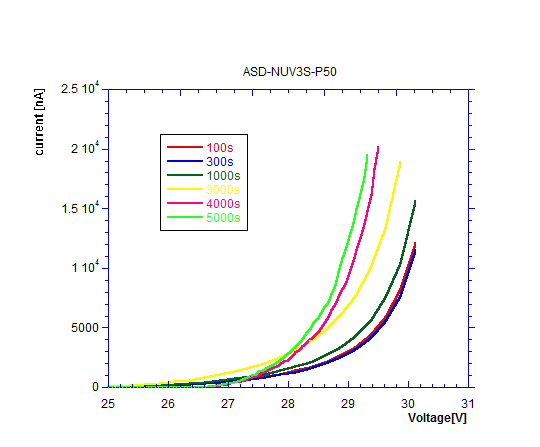}
\caption{\label{fig:IV} I-V curves of AdvanSiD ASD-NUV3S-P40 (top left), Hamamatsu S12572-050P (top right) and AdvanSiD ASD-NUV3S-P50 (bottom) models after each irradiation.}
\end{figure}

We recorded the I-V curves for all the devices before and shortly after each irradiation. 
Devices were irradiated once per day and measured two hours after the irradiation to allow for the decay of metastable nuclei.
In a few cases measurements were performed from few minutes to several hours after the irradiation to evaluate the contribution from metastable nuclei but no effect was detected.
For all devices the dark currents increase as the integrated doses increase. For example in Fig.~\ref{fig:IV}, the I-V curves for various neutron doses for Hamamatsu and AdvanSiD ASD-NUV3S models are shown.

\begin{figure}[htbp]
\centering % \begin{center}/\end{center} takes some additional vertical space
\includegraphics[width=0.49\textwidth]{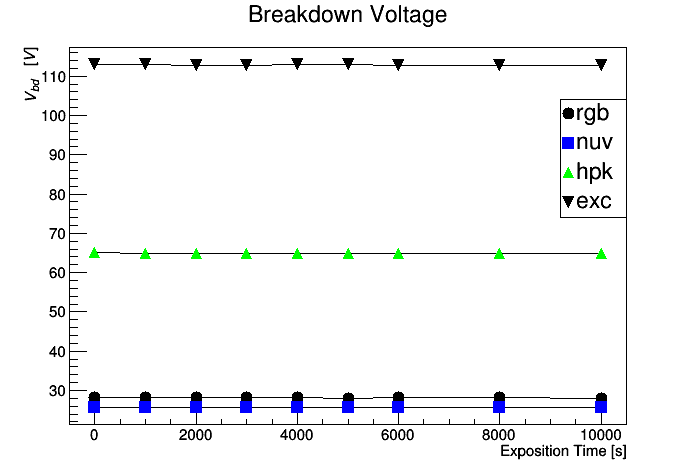}
\includegraphics[width=0.49\textwidth]{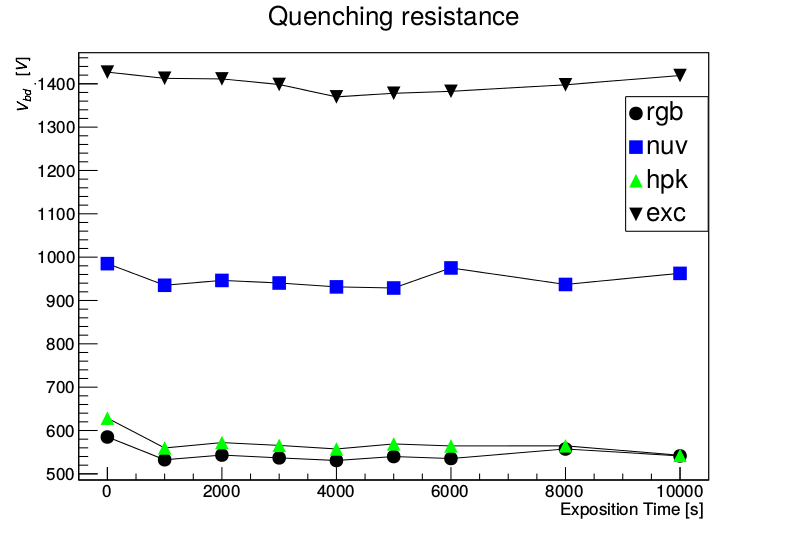}
\caption{\label{fig:compari} Breakdown voltage (left) and quenching resistance (right) as a function of irradiation time (rgb: AdvanSiD ASDRGB3S-P50, nuv: ASD-NUV3S-P40, hpk: Hamamatsu S12572-050P, exc: Excelitas C30742-33-050-X) .}
\end{figure}

The breakdown voltage is calculated with the Inverse Logarithmic Derivative (ILD) method defined with the following algorithm \cite{Chmill:2016msk}: %add reference https://arxiv.org/pdf/1605.01692.pdf
\begin{enumerate}
\item we record the inverse I-V curve from 0~V with steps of 0.05~V until the current reaches $20~\mu$A;
\item we calculate the logarithm of the curve;
\item we calculate the second derivative;
\item we define as breakdown voltage the maximum of the second derivative.
\end{enumerate}
For all the device under tests, the irradiation doses don't affect the breakdown voltages, as shown in Fig.~\ref{fig:compari} (left).

Also the quenching resistance of all devices is insensitive to the neutron dose (Fig.~\ref{fig:compari} right).

%The breakdown voltage does not change with neutron dose up to  $10^8$ neutron for all devices tested. The effect of neutron is remarkable viewing the current above the breakdown voltage. The current becomes higher with dose at the same operating voltage as shown in figure~\ref{fig:Vov}. This effect can be explained as a rise of noise due to neutron. This is an important limitation as you can’t increase the operating voltage.

Possible effects of annealing have not been studied systematically. The plan for the next future is to monitor constantly the SiPM operation.

\subsection{Noise evaluation}
\begin{figure}[htbp]
\centering % \begin{center}/\end{center} takes some additional vertical space
\includegraphics[width=0.49\textwidth]{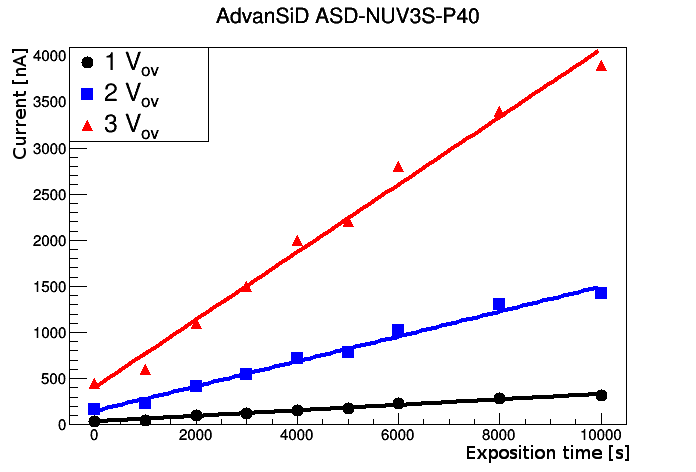}
\includegraphics[width=0.49\textwidth]{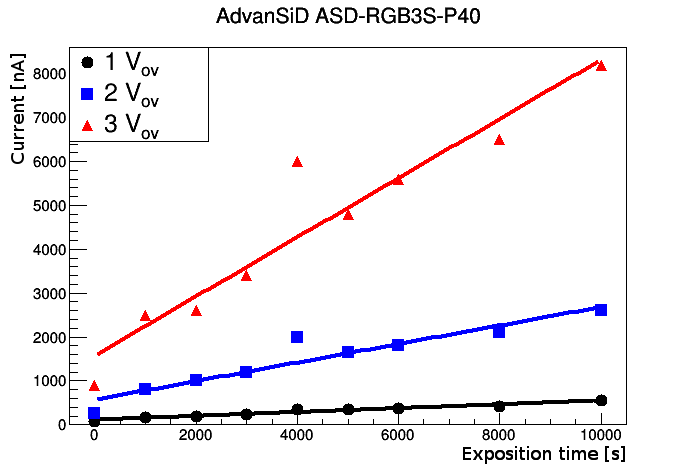}
\includegraphics[width=0.49\textwidth]{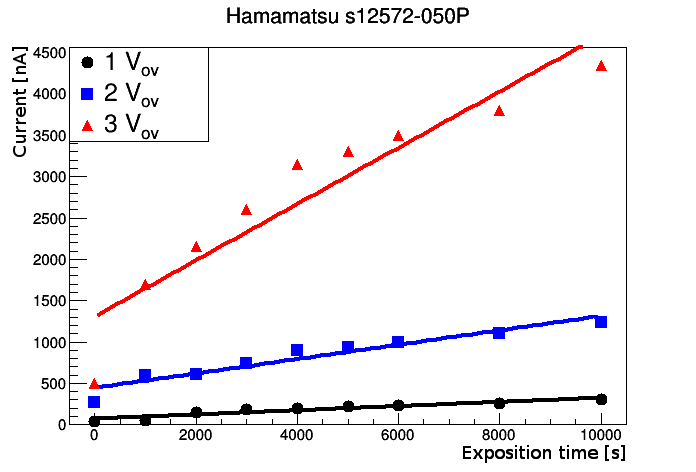}
\includegraphics[width=0.49\textwidth]{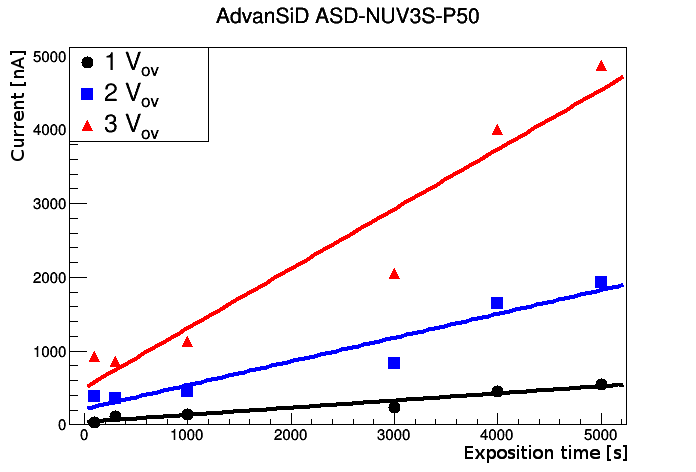}
\caption{\label{fig:Vov} Current at fixed over-voltage $V_{OV}$ as a function of the dose.}
\end{figure}

To evaluate the contribution of irradiation to the dark noise, we recorded at fixed over-voltage $V_{OV}$ the current as a function of the different doses. In Fig.~\ref{fig:Vov} the curves at $V_{OV}=1$~V, $V_{OV}=2$~V and $V_{OV}=3$~V for AdvanSiD ASD-NUV3S-P50, AdvanSiD ASD-NUV3S-P40, AdvanSiD ASD-RGB3S-P40 and Hamamatsu models are shown. For these devices the trend of the current is linear with respect the neutron dose for each value of over-voltage.

\section{Conclusions}
SiPMs have been irradiated with neutrons at doses comparable or larger to those expected during the data taking for the MEG~II experiment
to estimate the neutron induced radiation damage on their functionality.
We tested the SiPM model to be used in MEG~II (AdvanSiD ASD-NUV-P50), the most recent AdvanSiD devices (ASD-NUV3S-P40 and 
ASD-RGB3S-P40), a Hamamatsu (s12752-050P) and an Excelitas (C30742-33-050-X) with similar characteristics.

The most relevant effect of irradiation is the increase in dark current above the breakdown voltage. The measurements show a gradual increase. 
For all SiPM models, the increase of the current is proportional to the integrated doses, although in the case of the Excelitas model, 
because of technical and mechanical problems, we were not able to measure the current with sufficient precision. 

When considering the irradiation time delivering a fluence comparable to the total fluence expected in MEG~II $(3-5)\times 10^3$s, 
the main effect on the SiPM employed in MEGG~II (AdvanSiD ASD-NUV-P50), as visible in Fig.~\ref{fig:Vov} in the bottom, right panel, 
is an increase in dark current at $V_{OV}=3$~V up to $\sim 5\,\mu$A. This increase is not expected to influence significantly the
timing resolution of the devices during the experiment.

\newpage
\bibliographystyle{JHEP}

\begin{thebibliography}{99}

\bibitem{Adam:2013vqa} Adam, J. and others,
      The MEG detector for ${\mu}^+\to e^+{\gamma}$ decay search,
      Eur. Phys. J C73 (4) 2365, (2013) arXiv:1303.2348 [physics.ins-det]

\bibitem{Baldini:2013ke} Baldini, A. and others, MEG Upgrade Proposal, 2013,
      arXiv:1301.7225 [physics.ins-det]

\bibitem{barbieri} Barbieri, R. and Hall, L. and Strumia, A., 
Violations of lepton flavour and {CP} in super symmetric unified theories,
Nucl. Phys. B 445, (2--3), 219, (1995), arXiv:hep-ph/9408406

\bibitem{hisano-1999} Hisano, J. and Kurosawa, K. and Nomura, Y.,
     Large squark and slepton masses for the first-two generations in the  anomalous U(1) SUSY breaking models,
     Phys. lett. B 445, 316 (1999), arXiv:hep-ph/9810411

\bibitem{calibbi} Lepton flavor violation from supersymmetric grand unified theories: 
  where do we stand for MEG, PRISM/PRIME, and a super flavor factory, 
  Phys. Rev. D 74, 116002 (2006), arXiv:hep-ph/0605139

\bibitem{megtc} M. De Gerone {\it et al.},
  Development and commissioning of the Timing Counter for the MEG Experiment,
  IEEE Trans. on Nucl. Sci. {\bf Vol.59, No.2}, (2012) 379-388
  doi:10.1109/TNS.2012.2187311,
  arXiv:1112.0110 [physics.ins-det]

\bibitem{megtc1} M.~De Gerone {\it et al.},
  Design and test of an extremely high resolution Timing Counter for the
  MEG II experiment: preliminary results,
  Jour.\ Inst.\ 9 (2014) C02035, doi:10.1088/1748-0221/9/02/C02035, arXiv:1312.0871 [physics.ins-det].

\bibitem{megtc2} M.~De~Gerone {\it et al.} [MEG TC Collaboration],
  A high resolution Timing Counter for the MEG II experiment,
  Nucl.\ Instr.\ \&\ Meth.\ A {\bf 824} (2016) 92-95, doi:10.1016/j.nima.2015.11.022

\bibitem{odd:sm1}
D. Alloni, A. Borio di Tigliole, J. Bruni, M. Cagnazzo, R. Cremonesi, G. Magrotti, M. Oddone, F. Panza, M. Prata, A. Salvini, ''Neutron flux characterization of the SM1 sub-critical multiplying complex of the Pavia University'', Progress in Nuclear Energy, Volume 67, August 2013, Pages 98-103, ISSN 0149-1970.

\bibitem{Chmill:2016msk} 
  V.~Chmill, E.~Garutti, R.~Klanner, M.~Nitschke and J.~Schwandt,
  ``Study of the breakdown voltage of SiPMs,''
  Nucl.\ Instrum.\ Meth.\ A {\bf 845}, 56 (2017)
  doi:10.1016/j.nima.2016.04.047
  [arXiv:1605.01692 [physics.ins-det]].

\bibitem{mcnp} https://mcnp.lanl.gov/

\bibitem{pube} Edward Anderson, M., Bond Jr., W.H., 
Neutron spectrum of a plutonium - beryllium source, 
Nuc. Phys. 43, 330 (1963)

\bibitem{sandII} http://www.rist.or.jp/rsicc/app/sand\_2.htm

%% Please avoid comments such as "For a review'', "For some examples",
%% "and references therein" or move them in the text. In general,
%% please leave only references in the bibliography and move all
%% accessory text in footnotes.
%
%% Also, please have only one work for each \bibitem.
%
\end{thebibliography}

\end{document}